\documentclass[twocolumn,seceq,letterpaper]{jpsj3}
\usepackage{graphicx}
\usepackage{txfonts}
\usepackage{amsmath, amssymb, bm}
\usepackage{braket}
\usepackage{color}          
\usepackage{comment}
\usepackage{manfnt}
\usepackage{lineno}
\usepackage{here}

\title{Phonon-Induced Zero-bias Currents in Solids}

\author{
\name{Masao \surname{Ogata}$^{1,2}$}\thanks{E-mail: ogata@hosi.phys.s.u-tokyo.ac.jp}
and Hidetoshi \surname{Fukuyama}$^3$}

\inst{
$^1$Department of Physics, University of Tokyo, Hongo, Bunkyo-ku, Tokyo 113-0033, Japan 
$^2$Hongo 7-3-1, Bunkyo-ku, Tokyo 113-0033, Japan
\\
$^2$Research Institute for Energy Efficient Technologies, National Institute of Advanced Industrial Science and Technology (AIST), Umezono, Tsukuba, Ibaraki 305-8568, Japan
\\
$^3$Research Institute for Science and Technology, Tokyo University of Science, Shinjuku-ku, Tokyo 162-8601, Japan
} 
\date{\today}

\abst{
Zero-bias current induced by injected phonons in metals and one-dimensional 
charge density wave (CDW) systems attached on the surface of the piezoelectric substrate 
is investigated microscopically based on the second order response theory.  
In contrast to the shift currents discovered by von Baltz and Kraut in which 
the zero-bias current is induced by AC electric field in systems without  
inversion symmetry, propagating phonons break the inversion symmetry in the presesnt case. 
The effects of both deformation potential and piezoelectric potential are taken into account.
In the CDW system, zero-bias current appears below the transition 
temperature and its magnitude strongly depends on the position of the chemical potential. 
Possible experimental consequences are discussed.
}

\begin{document}

\maketitle

\gdef\QR{\bm Q \cdot \bm r}
\gdef\eP{\varepsilon_+}
\gdef\eO{\varepsilon_{+\Omega}}

\section{Introduction}

It is well-established that the existence of persistent current 
(zero-bias current) is a particular feature of superconductivity. 
Possibility of similar zero-bias currents has been predicted in insulators by von Baltz and Kraut 
in 1981 for systems without inversion symmetry under AC electric field\cite{vBK}
followed by many experiments confirming the prediction. 
This important finding by von Baltz and Kraut of combined effects of second order 
external perturbations and inversion symmetry breaking 
has versatile consequences in condensed matter science\cite{Review}.
Experimentally, the zero-bias current should be observed in the current ($I$)-voltage($V$) curve 
as shown in Fig.~1.
It is to be noted that the zero-bias current in the present case 
is a dissipative current induced by external forces, in contrast to the non-dissipative persistent current in superconductivity.

The acoustoelectric effect is an effect in which zero-bias DC currents are induced by injected  
bulk acoustic phonons. 
This was first proposed by Parmenter\cite{Parmenter}, 
followed by some early experiments\cite{Weinreich1,Sasaki,Wang,Beale}.
In these cases, the zero-bias current flows even if the system has inversion symmetry 
because the finite propagating vector $\bm Q$ of the acoustic phonon breaks the inversion symmetry.
This acoustoelectric effect was analyzed in phenomenological 
theories\cite{WeinreichTh,Hutson,Ingebrigtsen1,Ingebrigtsen2,Wixforth},
in which the attenuation of phonons results in electric current 
through the momentum transfer from phonons to electrons. 
However, in the presence of scattering of electrons by disorder, it is not clear how these phenomenological 
theories based on the momentum conservation law are valid. 
It is thus necessary to develop a microscopic theory using the second order nonlinear response. 

Recently, acoustoelectric effects are studied experimentally in terms of surface acoustic waves (SAWs) 
or acoustic phonons propagating on the surface of a piezoelectric substrate. 
A zero-bias current is observed in graphene placed on 
LiNbO$_3$\cite{Miseikis,Bandhu1,Bandhu2,Bandhu3,Zheng,Hernandez,Lane,Zhao,Mou}. 
Theoretically, Bhalla et al.\cite{Vignale} developed a microscopic theory based on thermal Green's functions
to study the case of graphene assuming that the SAW gives rise to the deformation potential, 
piezoelectric potential, and sound-induced vector potential\cite{Suzuura,ReviewGra}
on the two-dimensional electron system of graphene.
[It is to be noted that the sound-induced vector potential is specific 
to the honeycomb lattice\cite{Suzuura}.]
In the present paper, based on the microscopic method of Bhalla et al, 
we study the effects of SAW in metals in general. 
In particular, it is found by detailed analysis that this zero-bias current will have pronounced effects 
in the presence of CDW.

In section 2, we discuss the microscopic theory of the zero-bias current 
in terms of thermal Green's functions based on the second order response theory
and apply the method to one-, two-, and three-dimensional metals. 
As the interaction between electron and phonon, we consider deformation potential and 
piezoelectric potential caused by the injected SAW.
The obtained results show that the induced zero-bias current is proportional to the 
density of states multiplied by the momentum derivative of group velocity.
In Section 3, we study the zero-bias current in one-dimensional CDW systems 
based on mean-field Hamiltonian. 
It is shown that the zero-bias current is induced below the transition temperature 
and its magnitude depends on the position of the chemical potential. 
In Section 4, we discuss possible experimental consequences. 

\begin{figure}[]
\includegraphics[width=10cm]{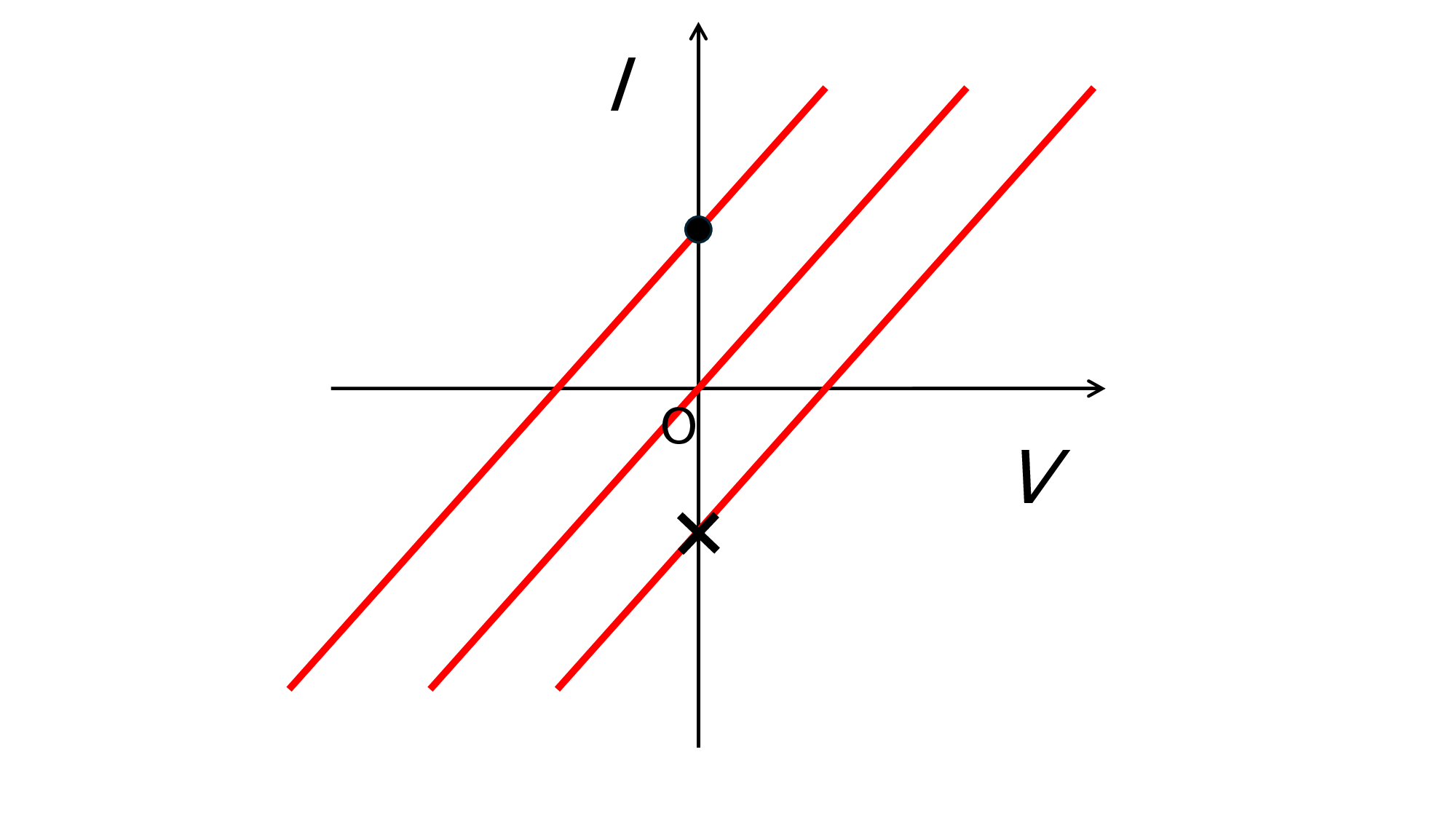}
\caption{(Color online)
Schemetic representation of $I$(current)-$V$(voltage) curves 
in the presence of zero-bias currents together with ordinary Ohmic currents. 
Zero-bias current can be either plus (solid circle) for holes or minus (cross) for electrons
at $V=0$.
$V$ represents the voltage applied to the sample. 
}
\label{fig:Zero-bias}
\end{figure}

\section{Zero-bias current in metals}

In this paper, we study electron systems 
interacting with an external propagating longitudinal acoustic (LA) phonon on the surface of substrate given by 
\begin{equation}
\bm u (\bm r, t) = {\bm u}_0 \cos (\bm Q \cdot \bm r - \Omega t)
= \frac{{\bm u}_0}{2} \left ( e^{i\bm Q \cdot \bm r} e^{-i\Omega t} + e^{-i\bm Q \cdot \bm r} e^{i\Omega t} \right),
\label{eq:Phonon}
\end{equation}
where we assume that the amplitude ${\bm u}_0$ as a classical external field. 
%
%
First, we consider the situation in which the long-wavelength LA phonon induces local variation of
average density of ions 
\begin{equation}
\rho_{\rm i} (\bm r) = \frac{\rho_{\rm i}}{1+ {\rm div}\ {\bm u}(\bm r, t)} 
\sim \rho_{\rm i}(1- {\rm div}\ \bm u (\bm r, t)), 
\end{equation}
which leads to an external interaction Hamiltonian, $H'$, as in the electron-phonon interaction in metals.
In addition, when the substrate is piezoelectric material, there is 
a piezoelectric potential\cite{Vignale,MahanBook}. 
Considering these two interactions, the external interaction Hamiltonian is given by
\begin{equation}
H'(t) = \sum_{\bm k, \sigma} [ 
c_{\bm k + \bm Q, \sigma}^\dagger A_{\bm Q} c_{\bm k, \sigma}^{\phantom{\dagger}} e^{-i\Omega t} + 
c_{\bm k - \bm Q, \sigma}^\dagger A_{-\bm Q} c_{\bm k, \sigma}^{\phantom{\dagger}} e^{i\Omega t}],
\label{eq:ElPhonon}
\end{equation}
where $c_{\bm k, \sigma}$ and $c_{\bm k, \sigma}^\dagger$ are the electron annihilation and creation operator
and $A_{\bm Q}$ is defined as
\begin{equation}
A_{\bm Q} = \frac{i g_1}{2} \bm Q \cdot {\bm u}_0+\frac{1}{2} g_{\rm P} u_0.
\label{eq:El-Phonon0}
\end{equation}
Here, $g_1$ is the deformation potential and $g_{\rm P}$ represents the 
coupling constant for the piezoelectric potential.
Note that the two effects can be treated together in the form of
eq.~(\ref{eq:El-Phonon0})\cite{Vignale,MahanBook}. 

\begin{figure}[t]
\hskip 1truecm
\includegraphics[scale=0.35]{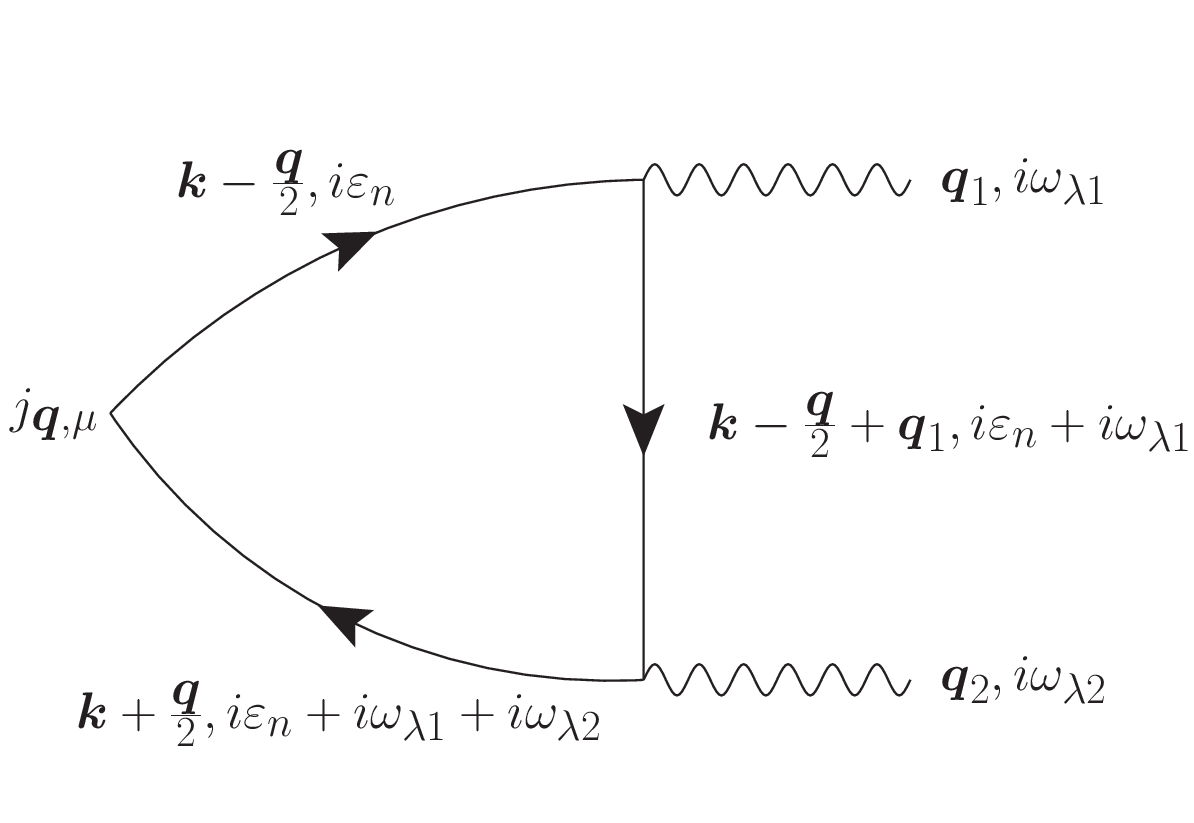}
\caption{Feynman diagram for the second order response theory. 
Solid lines represent thermal Green's functions and the wavy lines represent the perturbation $H'(t)$. 
The relation $\bm q = \bm q_1 + \bm q_2$ has been noted. 
Note that the vertices associated with the perturbation have Matsubara frequencies 
$i\omega_{\lambda 1}$ and $i\omega_{\lambda 2}$ with $\omega_{\lambda 1}>0$ 
and $\omega_{\lambda 2}>0$.
}
\label{fig:Feynman}
\end{figure}

In the second order response theory with respect to the electron-phonon interaction $H'(t)$ in eq.~(\ref{eq:ElPhonon}),
the current density in the $\mu$-direction with momentum $\bm q$ 
\begin{equation} 
j_{{\bm q},\mu} = - |e| \sum_{\bm k, \sigma} c_{\bm k-\frac{\bm q}{2}, \sigma}^\dagger v_{{\bm k},\mu}
c^{\phantom{\dagger}}_{\bm k+\frac{\bm q}{2}, \sigma},
\end{equation}
is obtained from the analytic continuation of a response function
\begin{equation}\begin{split}
&\Phi^{(2)}_{\bm q}(i\omega_{\lambda 1}, i\omega_{\lambda 2}) \cr
&=
- |e| k_{\rm B}T \sum_{n} \sum_{\bm k, \sigma} {\rm Tr}\biggl[ 
v_{\bm k, \mu} \mathcal G \left(\bm k-\frac{\bm q}{2}+\bm q_1 + \bm q_2, 
i\varepsilon_n + i\omega_{\lambda 1} + i\omega_{\lambda 2} \right)\cr
&\times A_{\bm q_2} 
\mathcal G \left( k - \frac{\bm q}{2} + \bm q_1, i\varepsilon_n + i\omega_{\lambda 1} \right) 
A_{\bm q_1}\ \mathcal G \left(\bm k - \frac{\bm q}{2}, i\varepsilon_n \right) \biggr] \delta_{\bm q - \bm q_1- \bm q_2, \bm 0},
\label{eq:GeneralFormXX}
\end{split}\end{equation}
where $v_{\bm k, \mu}$ is the electron group velocity defined as 
$v_{\bm k,\mu} = \partial H_{\bm k}/\hbar \partial k_\mu$,
$\mathcal G (\bm k, i\varepsilon_n) = [i\varepsilon_n +\mu - H_{\bm k}]^{-1}$ is the thermal Green's function of electrons, 
$k_{\rm B}$ is the Boltzmann constant, and the trace (Tr) is taken when the model has a matrix form.
Figure \ref{fig:Feynman} shows the Feynman diagram of the corresponding process where the solid lines represent the
Green's function, $\mathcal G$, and the wavy lines represent the perturbation $H'(t)$. 
The same process has been studied by Bhalla et al for graphene\cite{Vignale}. 
Note that the vertices associated with the perturbation have Matsubara frequency 
$i\omega_{\lambda 1}$ and $i\omega_{\lambda 2}$ with $\omega_{\lambda 1}>0$ 
and $\omega_{\lambda 2}>0$, and the analytic continuation is taken as 
$i\omega_{\lambda 1} \rightarrow \hbar\Omega_1+i\delta$ and 
$i\omega_{\lambda 2} \rightarrow \hbar\Omega_2+i\delta$ irrespective of the signs of 
$\Omega_1$ and $\Omega_2$ of the external fields.\cite{Kubo,FW,Vignale}
This is because the perturbation by $H'(t)$ is not internal but external and 
should be treated as adiabatic as in the case of shift current\cite{vBK,Sipe}.

For $\bm q_1, \bm q_2, \omega_{\lambda1}$, and $\omega_{\lambda 2}$ in Fig.~\ref{fig:Feynman}, we 
consider the two cases
\begin{equation}\begin{split}
{\rm (a)}\quad &\bm q_1 = \bm Q, i\omega_{\lambda 1} \rightarrow \hbar \Omega + i\delta, 
\bm q_2 = -\bm Q, i\omega_{\lambda 2} \rightarrow -\hbar\Omega + i\delta, \cr
{\rm (b)}\quad &\bm q_1 = -\bm Q, i\omega_{\lambda 1} \rightarrow -\hbar\Omega + i\delta, 
\bm q_2 = \bm Q, i\omega_{\lambda 2} \rightarrow \hbar\Omega + i\delta, 
\end{split}\end{equation}
%
which lead to the expectation value of uniform DC current $\langle j_{\bm q=\bm 0, x}\rangle$ 
as follows after summation over the fermion Matsubara frequency $i\varepsilon_n$.
\begin{equation}\begin{split}
\langle j_x \rangle &= 2|e| \sum_{\bm k}
\int \frac{d\varepsilon}{\pi} f(\varepsilon)  {\rm Tr} \biggl[ v_{\bm k, x} \cr
&\times \biggl\{ 
G_R(\bm k, \varepsilon) A_{-\bm Q} G_R(\bm k + \bm Q, \varepsilon+\hbar\Omega)
A_{\bm Q} \left[ {\rm Im} G_R(\bm k, \varepsilon) \right] \cr
&+G_R(\bm k, \varepsilon - \hbar\Omega) A_{-\bm Q} 
\left[ {\rm Im} G_R(\bm k + \bm Q, \varepsilon) \right]
A_{\bm Q} G_A(\bm k, \varepsilon-\hbar\Omega)  \cr
&+\left[ {\rm Im} G_R(\bm k, \varepsilon) \right] A_{-\bm Q} 
G_A(\bm k + \bm Q, \varepsilon+\hbar\Omega) A_{\bm Q} G_A(\bm k, \varepsilon) \cr
&+ ( \bm Q \rightarrow -\bm Q, \Omega \rightarrow -\Omega) \biggr\} \biggr],
\label{eq:J000}
\end{split}\end{equation}
where $f(\varepsilon)$ is the Fermi distribution function defined as $f(\varepsilon) = 1/[e^{(\varepsilon-\mu)/k_{\rm B}T}+1]$,
the factor $2$ comes from the spin summation, and $(\bm Q\rightarrow -\bm Q, \Omega \rightarrow -\Omega)$ means 
the contribution with $\bm Q$ and $\Omega$ in the preceding terms being replaced by $-\bm Q$ and $-\Omega$. 
In eq.~(\ref{eq:J000}), 
$G_R (\bm k, \varepsilon)$ is the retarded Green's function, 
$G_A (\bm k, \varepsilon) = G_R^\dagger (\bm k, \varepsilon)$ is the advanced Green's function, and 
${\rm Im} G_R(\bm k, \varepsilon) = (G_R(\bm k, \varepsilon)-G_A(\bm k, \varepsilon))/2i$. 

To evaluate $\langle j_x \rangle$ at finite $\bm Q$, we note that 
expansion with respect to $\bm Q$ is allowed since the wave vector 
$|\bm Q|$ of LA phonon is small compared with the characteristic wave vector of electrons. 
Therefore, we note
\begin{equation}\begin{split}
G_{R(A)} (\bm k + \bm Q, \varepsilon) &= G_{R(A)} (\bm k, \varepsilon) +  
G_{R(A)} (\bm k, \varepsilon) \bm Q \cdot \bm \gamma_{\bm k} G_{R(A)} (\bm k, \varepsilon) \cr
&+ O(|\bm Q|^2),
\end{split}\end{equation}
where $\bm \gamma_{\bm k}$ is defined as $ \bm \gamma_{\bm k} =\hbar \bm v_{\bm k}=\partial H_{\bm k}/\partial \bm k$,
and we have used the relationship 
$\partial G_{R(A)} /\partial \bm k = G_{R(A)} \bm \gamma_{\bm k} G_{R(A)}$ that holds also in the 
case of matrix form. 
Furthermore, since the phonon frequency $\Omega=v_s |\bm Q|$ ($v_s$ is the sound velocity) 
is small compared with electronic energy, $\langle j_x \rangle$ is given by the expansion with respect to 
$\Omega$. 
We find that $\langle j_x \rangle$ vanishes when $\Omega=0$. Therefore, in the lowest order of $\Omega$, we obtain
\begin{equation}\begin{split}
\langle j_x \rangle
%
%
=& 2|e| \Omega 
\int \frac{d\varepsilon}{\pi} f'(\varepsilon) \sum_{\bm k} Q {\rm Im}\ {\rm Tr} \cr
&\biggl[       \gamma_{\bm k, x} G_R A_{-\bm Q} G_R \gamma_{\bm k, x} G_R A_{\bm Q} G_A  \cr
&-\frac{1}{2} \gamma_{\bm k, x} G_R A_{-\bm Q} G_R \gamma_{\bm k, x} G_R A_{\bm Q} G_R 
+ (\bm Q\rightarrow -\bm Q) \biggr],
\label{eq:FinalJ}
\end{split}\end{equation}
where we have used $\bm Q=(Q,0,0)$ and $v_{\bm k, x} = \gamma_{\bm k, x}/\hbar$. 
In eq.~(\ref{eq:FinalJ}), $G_{R(A)}$ is the abbreviation of $G_{R(A)} (\bm k, \varepsilon)$ and $(\bm Q\rightarrow -\bm Q)$ 
means the contribution with $\bm Q$ in the preceding two terms being replaced by $-\bm Q$.

For the explicit summation over $\bm k$, we assume free electrons with 
isotropic effective mass $m$ in each dimension $d$ together with the 
damping $\Gamma$ independent of energy and momentum, 
$G_{R} (\bm k, \varepsilon) = 1/[ \varepsilon+i\Gamma - \hbar^2 k^2/2m]$, 
and $G_A=G_R^*$.
Here we assume that electronic damping $\Gamma$ large enough 
compared to phonon energy, i.e. $\Gamma > \hbar\Omega = \hbar v_s |{\bm Q}|$ 
($\cong 10^{-6}$ eV in cases of interest).
Then summations over $\bm k$ in eq.~(\ref{eq:FinalJ}) 
in each dimension can be performed 
as follows,
\begin{equation}\begin{split}
&d=1, \cr & \frac{m^2 L}{4\hbar^4}{\rm Re}  
\left[ -\frac{1}{\kappa_R^5} + \frac{i\hbar^2}{m\Gamma \kappa_R^3}
+\frac{\hbar^4}{m^2\Gamma^2 \kappa_R} + \frac{i\hbar^6}{2m^3\Gamma^3} 
(\kappa_R-\kappa_A) \right], 
\label{eq:Final1D}
\end{split}\end{equation}
\begin{equation}\begin{split}
&d=2, \cr & \frac{m^2 L^2}{6\pi \hbar^4} {\rm Im}
\left[ -\frac{1}{\kappa_R^4} + \frac{3i\hbar^2}{2m\Gamma \kappa_R^2} 
+ \frac{3i\hbar^4}{4m^2 \Gamma^2} 
\left(\frac{\pi}{2}+\tan^{-1} \frac{\varepsilon}{\Gamma} \right) \right], 
\label{eq:Final2D}
\end{split}\end{equation}
\begin{equation}\begin{split}
&d=3, \cr & \frac{m^2 L^3}{24\pi \hbar^4} {\rm Re}  
\left[ \frac{1}{\kappa_R^3} - \frac{3i\hbar^2}{m\Gamma \kappa_R}
+\frac{3\hbar^4 \kappa_R}{m^2\Gamma^2} +\frac{i\hbar^6}{2m^3\Gamma^3} 
(\kappa_R^3-\kappa_A^3) \right], 
\label{eq:Final3D}
\end{split}\end{equation}
where
$\kappa_R = \sqrt{2m(\varepsilon+i\Gamma)}/\hbar$ (Im $\kappa_R>0$) and 
$\kappa_A = \sqrt{2m(\varepsilon-i\Gamma)}/\hbar$ (Im $\kappa_A>0$). 
Note that $\kappa_A = - \kappa_R^*$, so that 
$\kappa_R-\kappa_A= \kappa_R+\kappa_R^*$ and 
$\kappa_R^3-\kappa_A^3 = \kappa_R^3+(\kappa_R^*)^3$ are both real and  
the last terms in eqs.~(\ref{eq:Final1D}) and (\ref{eq:Final3D}) do not contribute.

For $\Gamma << \varepsilon_{\rm F}$, we see that the dominant 
contribution in $\langle j_x \rangle$ is in the order of 
$\varepsilon_{\rm F}^2/\Gamma^2$ at low temperatures where 
$f'(\varepsilon) \sim -\delta(\varepsilon-\varepsilon_{\rm F})$ holds. In this case, we obtain
\begin{equation}\begin{split}
d=1, \qquad \langle j_x \rangle &= -|e|
\frac{Q \Omega \hbar |A_{\bm Q}|^2 L}{\pi \Gamma^2 \sqrt{2m\varepsilon_{\rm F}}}, \cr
d=2, \qquad \langle j_x \rangle &= -|e| 
\frac{Q \Omega |A_{\bm Q}|^2 L^2}{2\pi \Gamma^2}, \cr
d=3, \qquad \langle j_x \rangle &= -|e|
\frac{Q \Omega |A_{\bm Q}|^2 L^3 \sqrt{2m\varepsilon_{\rm F}}}{2\pi^2 \hbar \Gamma^2}, 
\label{eq:DominantJ}
\end{split}\end{equation}
where $|A_{\bm Q}|^2=u_0^2 (g_1^2 Q^2 + g_{\rm P}^2)/4$.
We see that $\langle j_x \rangle$ is proportional to the density of 
states in each dimension multiplied by the $k_x$-derivative of the group velocity $v_{\bm k}$.
An alternative derivation of eq.~(\ref{eq:DominantJ}) is given in Appendix. 
The present result shows that the phonon-induced zero-bias currents appear even in 
simple metals.

\section{Zero-bias current in one-dimensional CDW system}

In the following, we study the case of a one-dimensional CDW system described by the Hamiltonian
\begin{equation}
H_{\bm k} = \xi_k \sigma_z+ \Delta \sigma_x,
\label{eq:CDWHamiltonian}
\end{equation}
interacting with external propagating LA phonon on the surface of substrate, as shown in eq.~(\ref{eq:Phonon}). 
Here, $\sigma_x$ and $\sigma_z$ are the $2\times 2$ Pauli matrices, 
$\xi_k= v_{\rm F} \hbar k$ with $v_{\rm F}$ being the Fermi velocity, and $\Delta$ represents the CDW order parameter. 
For the right-going (left-going) electrons, the momentum $k$ is measured from 
$k_{\rm F}$ ($-k_{\rm F}$). 
In this case, $\gamma_{\bm k, x}$ is a matrix $\gamma_{\bm k, x}=v_{\rm F} \hbar \sigma_z$ and 
the Green's functions are 
\begin{equation}
G_{R} = \frac{\eP + \xi_k \sigma_z + \Delta \sigma_x}{D_R}, \qquad
G_{A} = \frac{\varepsilon_- + \xi_k \sigma_z + \Delta \sigma_x}{D_A}, 
\end{equation}
where $\varepsilon_\pm = \varepsilon \pm i\Gamma$,  $D_R = \eP^2- \xi_k^2-\Delta^2$ and $D_A = (D_R)^*$.
Taking the trace in eq.~(\ref{eq:FinalJ}), we obtain
\begin{equation}\begin{split}
\langle j_x \rangle &= 4|e| Q \Omega v_{\rm F}^2 \hbar^2 |A_{\bm Q}|^2 \int \frac{d\varepsilon}{\pi} f'(\varepsilon) 
\sum_{k} \cr 
&\times {\rm Im} \biggl[ \frac{2}{D_R D_A} \left\{ 1-\frac{2i\Gamma \eP}{D_R} 
+\frac{8\xi_k^2}{D_R^2} \varepsilon \eP \right\} \cr
&-\frac{1}{D_R^2} \left\{ 1+ \frac{8\xi_k^2}{D_R^2} \eP^2 \right\} \biggr].
\end{split}\end{equation}
The summation over $k$ results in 
\begin{equation}\begin{split}
\langle j_x \rangle &= -|e| Q \Omega v_{\rm F} \hbar |A_{\bm Q}|^2 L
\int \frac{d\varepsilon}{\pi} \left( -f'(\varepsilon) \right) \cr
&\times {\rm Re} \biggl[ \frac{\Delta^2}{\varepsilon^2 \Gamma^2 x_+} + \frac{i\Delta^2}{\varepsilon \Gamma x_+^3} 
- \frac{\Delta^2}{x_+^5} \biggr],
\label{eq:CDWfinal}
\end{split}\end{equation}
with $x_+=\sqrt{\eP^2-\Delta^2}$ (${\rm Im}x_+>0$). 
It is to be noted that the zero-bias current vanishes in the normal state with $\Delta=0$. 
This seems to contradict the results in the previous section, where we have shown that
the simple metals have zero-bias currents. 
However, it is not a contradiction because the present CDW model is special in the sense that 
the group velocity $v_{\bm k, x}$ is constant ($v_{\bm k, x} = \pm v_{\rm F}$) 
and thus its $k_x$-derivative vanishes. 
%
\begin{figure}[t]
\includegraphics[scale=0.65]{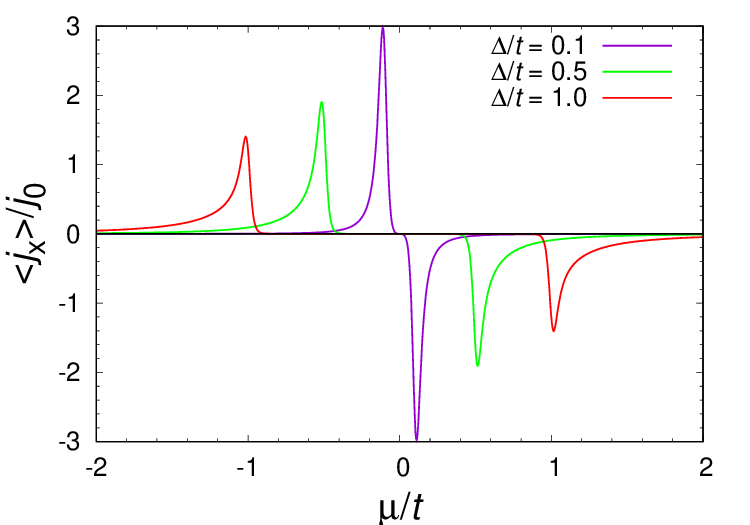}
\caption{$\langle j_x \rangle / j_{0}$ as a function of $\mu/t$ at $T=0$, where 
$j_{0} = |e| Q \Omega v_{\rm F} \hbar |A_{\bm Q}|^2 L/(t\Gamma^2)$ 
with a unit of energy $t$. 
$\Gamma$ is fixed at $\Gamma/t = 0.05$ and $\Delta$ are chosen as $\Delta/t=0.1, 0.5$ and $1.0$. }
\label{fig:CDW}
\end{figure}

The $\mu$-dependence of $\langle j_x \rangle$ at $T=0$ divided by 
$j_{0} = |e| Q \Omega v_{\rm F} \hbar |A_{\bm Q}|^2 L/(t\Gamma^2)$ is shown in Fig.~\ref{fig:CDW}
for $\Gamma/t=0.05$ and for several values of $\Delta/t$.
Here, we choose $t\equiv v_{\rm F} \hbar /a$ as a unit of energy with $a$ being the lattice constant. 
When $\mu$ is positive (negative), $\langle j_x \rangle<0$ $(\langle j_x \rangle>0)$, 
which means that the 
electrons (holes) are dragged by the surface acoustic phonon leading to a negative (positive) current.
As seen from Fig.~\ref{fig:CDW}, there are peaks at $\mu=\pm \Delta$ when $\Delta$ is finite.
%
In the case of $\Gamma << |\mu|-\Delta$, we see that the dominant 
contribution at $T=0$ comes from the first term in eq.~(\ref{eq:CDWfinal}) proportional to $\Gamma^{-2}$, which gives
\begin{equation}
\langle j_x \rangle = 
-\frac{j_{0} \Delta^2 t}{\pi \mu^2 \sqrt{\mu^2-\Delta^2}}
\theta(\mu^2-\Delta^2) {\rm sign}\mu. 
\label{eq:CDWlimit}
\end{equation}
%
This expression explains the behavior of $\langle j_x \rangle$ in Fig.~\ref{fig:CDW}, 
i.e., $\langle j_x \rangle \propto -1/\mu^3$, in the region $|\mu| >> \Delta$.
On the other hand, 
when $\mu=\pm \Delta$, the divergence in eq.~(\ref{eq:CDWlimit}) is suppressed 
by the presence of $\Gamma$. 
In this case, the dominant contribution at $\mu=\pm \Delta$ turns out to be 
\begin{equation}
\langle j_x \rangle = \mp \frac{7 t}{8\pi \sqrt{\Gamma \Delta}} j_{0},
\label{eq:PeakValues}
\end{equation}
%
where we have used 
$x_+ = \sqrt{\pm 2i\Gamma\Delta - \Gamma^2}$ when $\mu=\pm \Delta$, and
$x_+ \sim (\pm 1 +i ) \sqrt{\Gamma\Delta}$ in the limit of $\Gamma<<\Delta$.
The peak value of $\langle j_x \rangle$ at $\mu=\pm \Delta$ becomes smaller 
as $\Delta$ becomes larger as seen in Fig.~\ref{fig:CDW}, 
which is consistent with the $\Delta$-dependence 
of the right-hand side of eq.~(\ref{eq:PeakValues}). 

\begin{figure}[t]
\vskip 1.2truecm
\hskip -1truecm
\includegraphics[width=10.5cm]{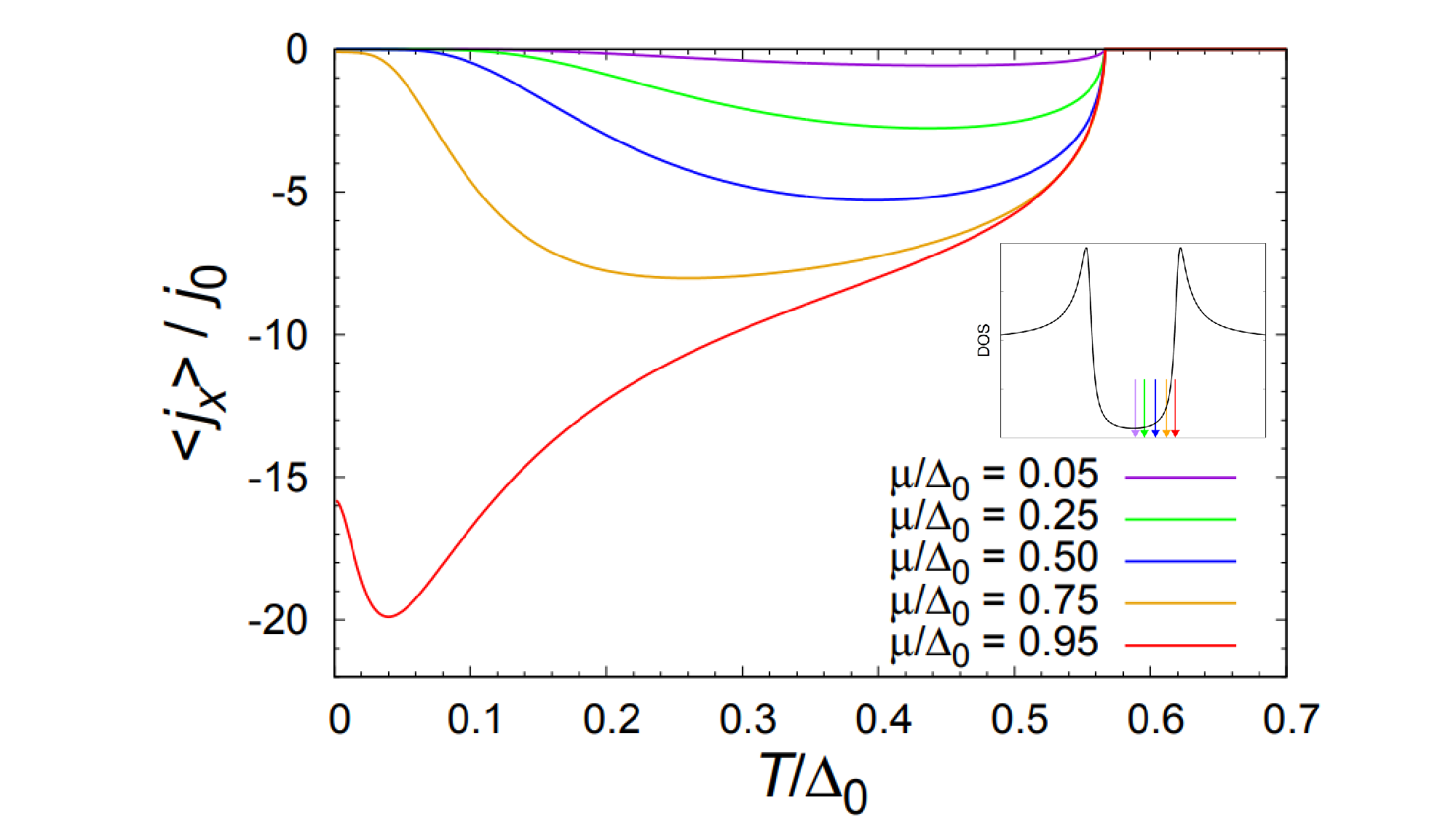}
\caption{$\langle j_x \rangle/j_0$ as a function of temperature 
for several values of the chemical potential $\mu$.
[$\mu/\Delta_0=0.05, 0.25, 0.50, 0.75$, and $0.95$ from top to bottom.] 
The unit of energy is taken as $\Delta_0$ that is the magnitude of the gap 
at $T = 0$. 
$\Gamma$ is fixed at $\Gamma/\Delta_0 = 0.10$.
The inset shows the positions of the chemical potentials. 
}
\label{fig:Temp}
\end{figure}

To discuss the temperature dependence of $\langle j_x \rangle$, we consider the mean-field theory 
of the CDW state with the Hamiltonian $H_k+\frac{K}{2}\Delta^2$, where the second term represents 
the elastic energy loss due to the CDW order parameter. 
The self-consistency equation for $\Delta$ becomes
\begin{equation}\begin{split}
K\Delta &= 2\sum_{k} \frac{\Delta}{E_k} \left( 1-2f(E_k) \right) \cr
&= \frac{1}{\pi v_{\rm F} \hbar} \int_{-t}^t d\xi \frac{\Delta}{E_\xi} \left( 1-2f(E_\xi) \right),
\label{SCeq}
\end{split}\end{equation}
where the factor 2 in the first line comes from the spin summation, in the second line 
$E_\xi=\sqrt{\xi^2+\Delta^2}$, and a cut-off energy is assumed to be $\pm t$.
Instead of solving eq.~(\ref{SCeq}),  we assume that the order parameter has a temperature dependence 
\begin{equation}
\Delta(T) = \Delta_0 \tanh \left[ \frac{\pi k_{\rm B} T_{\rm c}}{\Delta_0}
\sqrt{\frac{8}{7\zeta(3)}\ \frac{T_{\rm c}-T}{T}} \right],
\label{eq:DeltaT}
\end{equation}
between $0<T \le T_{\rm c}$, where $\zeta(3)$ is the Riemann zeta function,
$\Delta_0$ is the CDW order parameter at $T=0$
($\Delta_0 = 2t e^{-\pi v_{\rm F} \hbar K/2}$), 
and the phase transition temperature is $k_{\rm B}T_{\rm c} = \frac{e^\gamma}{\pi} \Delta_0 = 0.567 \Delta_0$
with $\gamma$ being Euler's constant ($\gamma \sim 0.577$). 
%
%
%
Using the form of $\Delta(T)$ in eq.~(\ref{eq:DeltaT}), we obtain temperature dependence of 
$\langle j_x \rangle$ by
\begin{equation}\begin{split}
\langle j_x \rangle &= -|e| Q \Omega v_{\rm F} \hbar |A_{\bm Q}|^2 L
\int \frac{d\varepsilon}{\pi} \left( -f'(\varepsilon) \right) \cr
&\times {\rm Re} \biggl[ \frac{\Delta(T)^2}{\varepsilon^2 \Gamma^2 x_+} + \frac{i\Delta(T)^2}{\varepsilon \Gamma x_+^3} 
- \frac{\Delta(T)^2}{x_+^5} \biggr].
\label{eq:CDWfinalTemp}
\end{split}\end{equation}
When we consider a single-band model described by the Hamiltonian (\ref{eq:CDWHamiltonian}),  
there is an electron-hole symmetry and the chemical potential is fixed at $\mu=0$. 
In this case, the zero-bias current is not induced since both the electrons and holes 
are equally dragged. 
However, in some materials, other overlapping bands exist near the Fermi energy
in addition to the one-dimensional CDW system. 
In such cases, the chemical potential deviates from $\mu=0$. 
When $\mu>0$ ($\mu<0$), the electrons (holes) are thermally excited more than 
the holes (electrons), leading to a negative (positive) induced current.
Taking account of this possibility of nonzero-$\mu$, we calculate the temperature dependence 
of $\langle j_x \rangle$ for several values of the chemical potential 
as shown in Fig.~\ref{fig:Temp} using the temperature dependence of $\Delta(T)$
with $\Delta_0=0.02t$. 
The inset of Fig.~\ref{fig:Temp} shows the positions of the chosen chemical potentials.
Here, we use $\Delta_0$ as a unit of energy and thus the phase transition temperature 
is $T_{\rm c} = 0.567 \Delta_0/k_{\rm B}$. 
When the temperature is lowered below $T_{\rm c}$, $\Delta(T)$ starts to develop and 
the induced current appears sharply. 
As the temperature decreases further, the gap $\Delta(T)$ increases while the derivative of the 
Fermi distribution function $(-f'(\varepsilon))$ becomes narrower as a function of $\varepsilon$. 
Combination of these two effects reduces the induced current at low temperatures.
When $\mu$ is close to zero, the absolute value of $\langle j_x \rangle$ is small 
because of the cancellation between the contributions from electrons and holes. 
In contrast, when $\mu$ is located close to the band edge ($\mu \sim \pm \Delta_0$),  
the absolute value of $\langle j_x \rangle$
remains large up to near zero temperature because $\langle j_x \rangle$ 
has sharp peaks at the edge of the gap as shown in Fig.~\ref{fig:CDW}.

\section{Possible experimental consequences and summary}


NbSe$_3$ is a candidate material of the one-dimensional CDW systems for observing 
the zero-bias current induced by SAW\cite{Kawada}. 
As indicated by the resistivity experiment\cite{Lear}, NbSe$_3$ shows metallic behavior 
even below the two successive phase transition temperatures ($T_1=140$K and $T_2=60$K) 
accompanied by enhancement of resistivity just below $T_1$ and $T_2$.
This means that there are other overlapping bands in addition to the CDW system. 
Therefore, the present results in Fig.~\ref{fig:Temp} will be helpful for the analysis of 
temperature dependence of zero-bias current observed experimentally.
Furthermore, from the sign of the zero-bias current, we obtain the information whether 
the carriers are electron-like or hole-like. 
If the sign is negative (positive), the carriers are 
electrons (holes) irrespective of the sign of $g_1$ and $g_{\rm P}$.

To understand the experimental results quantitatively, it is necessary to evaluate the values of 
the coupling constants $g_1$ and $g_{\rm P}$, although they are not known experimentally.  
Nevertheless, by studying the $Q$-dependence of induced current, 
we will be able to obtain the information whether the zero-bias current originates from the 
deformation potential or not,  
because the contribution from $g_1$ in $\langle j_x \rangle$ 
is proportional to $Q^3 \Omega=v_{\rm s} Q^4$, 
while that from $g_{\rm P}$ is proportional to $Q \Omega=v_{\rm s}Q^2$ 
as shown in eq.~(\ref{eq:CDWfinal}).
[Note that, in $|A_{\bm Q}|^2=u_0^2 (g_1^2 Q^2 + g_{\rm P}^2)/4$, 
the first term $g_1^2 Q^2$ comes from the deformation potential and the 
second term $g_{\rm P}^2$ comes from the piezoelectric potential.]

For a rough estimate of the magnitude of the zero-bias current, 
we assume that $u_0 g_{\rm P}=1$ meV and $u_0 g_1 Q=1$ meV tentatively.  
The other parameters are assumed to be $Q=6 \times 10^5$ m$^{-1}$, $v_{\rm F}=10^6$ m/sec, 
$\Gamma = 2$ meV, and $t=1$ eV.
The frequency of SAW is set to be 300 MHz, which corresponds to
$\Omega =1.88 \times 10^9$ sec$^{-1}$. 
When the cross-section of the CDW sample is 10$\mu$m $\times$ 0.04$\mu$m and 
the distance between the chains is 1 nm, 
we obtain $j_0\sim 10$nA. 

Let us discuss the relationship between our results 
of zero-bias current and those obtained phenomenologically. 
In the phenomenological theory, acoustoelectric 
current is given by\cite{WeinreichTh,Ingebrigtsen2,Wixforth}
\begin{equation}
j_{\rm AE} =-\frac{\alpha S \mu_{\rm e}}{v_s}, \label{eq:Phenomeno0}
\end{equation}
where $\alpha$ is the sound attenuation, $S$ is the power flux density of phonons, and 
$\mu_{\rm e}$ is the mobility of electrons.
Here, the complete momentum transfer from phonon to electrons is assumed. 
However, in realistic cases with disorder, such as impurities, 
phenomenological considerations based on momentum conservation are not valid, 
because the momentum transferred to the electron system will decay owing to the disorder.
Hence, there is no relationship between the phenomenological results and ours 
based on microscopic consideration 
under the assumption of $\hbar\Omega = \hbar v_s Q< \Gamma$.

To summarize, we have demonstrated that SAW induces zero-bias currents in metal in general. 
This is similar to the current discovered by von Baltz and Kraut for insulators 
without inversion symmetry under AC electric fields\cite{vBK}, 
in the sense that SAW plays roles of symmetry breaking in the present case of metals.
Present theoretical results can be tested by experiments especially in NbSe$_3$. 

\section*{Acknowledgments}
We thank very fruitful discussions with N.\ Nikaido, T.\ Kawada, M.\ Hayashi, 
K.\ Kitayama, and Y.\ Niimi. 
This work is supported by Grants-in-Aid for Scientific Research from the Japan Society 
for the Promotion of Science (No.\ JP23K03274 and No.\ JP23H01118), 
and JST-Mirai Program Grant (No.\ JPMJMI19A1).

\appendix 
\section{Zero-bias current of single-band models in the small $\Gamma$ limit}

In this appendix, we show that the zero-bias current $\langle j_x \rangle$ 
of single-band models is proportional to the density of states 
multiplied by the $k_x$-derivative of the group velocity $v_{\bm k}$ in the limit of 
$\Gamma << \varepsilon_{\rm F}$. 
In this limit, the dominant contribution to $\langle j_x \rangle$ comes from the 
first term in eq.~(\ref{eq:FinalJ}), which contains both the retarded ($G_R$) and 
advanced ($G_A$) Green's functions.
In the single-band models with energy dispersion $\varepsilon_{\bm k}$, 
\begin{equation}\begin{split}
\langle j_x \rangle =& -2i|e| Q\Omega \hbar^2 |A_{\bm Q}|^2 
\int \frac{d\varepsilon}{\pi} f'(\varepsilon) \sum_{\bm k}\cr
&v_{\bm k, x}^2 \biggl\{ 
\frac{1}{(\varepsilon-\varepsilon_{\bm k}+i\Gamma)^3 (\varepsilon-\varepsilon_{\bm k}-i\Gamma)} \cr
&-\frac{1}{(\varepsilon-\varepsilon_{\bm k}+i\Gamma) (\varepsilon-\varepsilon_{\bm k}-i\Gamma)^3} \biggr\},
\end{split}\end{equation}
where we have used $\gamma_{\bm k, x} = \hbar v_{\bm k, x}$ with 
$v_{\bm k, x} = \partial \varepsilon_{\bm k}/\hbar\partial k_x$. 
In the limit of $\Gamma << \varepsilon_{\rm F}$, the dominant contribution is obtained from the 
residues of the poles of the Green's functions\cite{Konye}. 
Therefore, we obtain
\begin{equation}
\langle j_x \rangle = 4|e| Q\Omega \hbar^2 |A_{\bm Q}|^2 \sum_{\bm k} 
\frac{v_{\bm k, x}^2 f''(\varepsilon_{\bm k})}{(2i\Gamma)^2} + O(\Gamma^0).
\end{equation}
Using the relation 
$\hbar v_{\bm k, x} f''(\varepsilon_{\bm k})=\partial f'(\varepsilon_{\bm k})/\partial k_x$,
we can perform the integration by part in the integral with respect to $\bm k$. 
As a result, the zero-bias current becomes
\begin{equation}
\langle j_x \rangle = |e| \frac{Q\Omega \hbar}{\Gamma^2} |A_{\bm Q}|^2 \sum_{\bm k} 
\frac{\partial v_{\bm k, x}}{\partial k_x} f'(\varepsilon_{\bm k}) + O(\Gamma^0).
\end{equation}
The right-hand side is proportional to the density of states 
multiplied by the $k_x$-derivative of the group velocity. 
In the simple model with $\varepsilon_{\bm k} = \hbar^2 k^2/2m$, we reproduce the results 
in eq.~(\ref{eq:DominantJ}).






\def\journal#1#2#3#4{#1 {\bf #2}, #3 (#4)}
\def\PR{Phys.\ Rev.}
\def\PRB{Phys.\ Rev.\ B}
\def\PRL{Phys.\ Rev.\ Lett.}
\def\JPSJ{J.\ Phys.\ Soc.\ Jpn.}
\def\PTP{Prog.\ Theor.\ Phys.}
\def\JPCS{J.\ Phys.\ Chem.\ Solids}
\def\JAP{J.\ Appl.\ Phys.}
\def\APL{Appl.\ Phys.\ Lett.}

\bibliographystyle{jpsj}
\bibliography{apssampNotes}

\end{document}